\documentclass[12pt]{article}
\newcommand{\zProof}{{\bf\underbar{Proof}.}\ }

\newcommand{\intT}{\int_0^T}
\newcommand{\zdia}{~~\rule{1mm}{2mm}\par\medskip}

\usepackage{color}
\usepackage{epsfig,cmmib57}
\usepackage{amssymb,amsmath,exscale}
\newcommand{\zaa}{\alpha}

\newcommand{\n}{\lambda_n}

\newcommand{\intt}{\int_0^t}
\newcommand{\ints}{\int_0^s}

\newtheorem{Theorem}{Theorem}
\newtheorem{Corollary}[Theorem]{Corollary}
\newtheorem{Lemma}[Theorem]{Lemma}

\newtheorem{Remark}[Theorem]{Remark}
\newtheorem{Definition}[Theorem]{Definition}

\newcommand{\ZIN}{\infty}
\newcommand{\zthe}{\theta}
\newcommand{\ZD}{\,{\rm d}\,}
\newcommand{\ZLA}{\label}

\newcommand{\zln}{\lambda_n^2}
\newcommand{\phin}{\phi_n}
\newcommand{\wn}{w_n}
\newcommand{\an}{(a-\zln)}
\newcommand{\mun}{\mu_n^2}
\author{
Andrei Halanay\thanks{Department of Mathematics and Informatics, University Politehnica of Bucharest, 313 Splaiul Independentei, 060042 Bucharest, Romania (halanay@mathem.pub.ro)}
\and L. Pandolfi\thanks{Dipartimento di Scienze Matematiche, Politecnico di Torino, Corso Duca degli Abruzzi 24---10129 Torino, Italy (luciano.pandolfi@polito.it)}
 }
\title{Lack of controllability of   thermal systems with memory  \footnote{The research of the first author is partially supported by Romanian CNCS Grant PN-II-ID-PCE-2011-3-0211. The research of the second author fits the  plans of INDAM-CNR and of the project ``Groupement de Recherche en Contr\^ole des EDP entre la France
et l'Italie (CONEDP)''. }}
\begin{document}
\maketitle

\begin{abstract}
Heat  equations with memory of Gurtin-Pipkin type (i.e. Eq.~(\ref{eq:CGM}) with $ \alpha=0 $) have controllability properties which strongly resemble  those of the wave equation. Instead, recent counterexamples show  that when $ \alpha>0 $ the control properties do not parallel those of the (memoryless) heat equation, in the sense that there are initial conditions in $ L^2(\Omega) $ which cannot be controlled to   zero. The proof of this fact consists in the construction of two quite special examples of systems with memory which cannot be controlled to zero. Here we prove that lack of controllability holds in general, for every smooth memory kernel $ M(t) $.
\end{abstract}

\section{Introduction}

The following integro-differential equation  is often used to model thermal systems with memory, see~\cite{ColemGURTIN,Preziosi}:
 
\begin{equation}
\ZLA{eq:CGM}
w _t=\alpha\Delta w  +\intt M(t-s)\Delta w (s)\ZD s\,,\qquad w (0)=\xi\,.
\end{equation}
Here $ w =w (x,t) $ and $ x\in\Omega $, a bounded region with  smooth   boundary (we require  of class $ C^1 $, and $ \Omega$ locally on one side of $ \partial\Omega $).

The time $ t=0 $ is the time after which a boundary control $ f $ is applied to the system,
\[ 
w (x,t)=f(x,t)\qquad x\in\Gamma= \partial\Omega\,,\quad t>0\,.
 \]
Note that we implicitly assume that the system is at rest for negative times,   $w(t)=0$   if $ t<0 $.

The number $ \alpha $ is nonnegative. If $ \alpha  $ is zero then    we get a model proposed by Gurtin and Pipkin in~\cite{GurtinPipkin}.  The controllability, when $ \alpha=0 $, has been studied in several paper, see references below. So, here we explicitly assume
\[ 
\alpha>0
 \]
 and we call Eq.~(\ref{eq:CGM}) the (CGM)   model (after   Colemann and Gurtin).

 It appears that (CGM) has been rarely studied from the control point of view. Our goal in this paper is to understand  whether the point $ \xi_0=0 $ can be hitted at time $ T>0 $, as it is the case for the memoryless heat equation, i.e. the special case of (CGM) obtained when $ M(t)\equiv 0 $.
 
 The precise definition of controllability requires that we specify the properties of the solutions. The following results are proved   in Section~\ref{sect:PreliDEFIN}, where the definition of ``solution'' can be found:
 
 \begin{Theorem}\ZLA{teo:propreSOL}
 Let $ M(t)\in C^1(0,+\infty) $.
For every $ f\in L^2(0,T;L^2(\Gamma)) $ and for every initial condition $ \xi\in L^2(\Omega) $ there exists a unique solution $ w (\cdot,T)=w ^{f,\xi}(\cdot,T)\in L^2(0,T;L^2(\Omega)) $.  
\end{Theorem}
The solution is not continuous in time, see the examples in~\cite[p.~217]   {Lions}, unless $ f(t) $ is smooth. So, pointwise computation of $ w (\cdot,t) $ in $ L^2(\Omega) $   is meaningless in general. However, let $ A $ be the operator in $ L^2(\Omega) $:
\begin{equation}
\ZLA{eq:defiOPERa}
{\rm dom}\,A=H^2(\Omega)\cap H_0^1(\Omega)\,,\qquad A\phi=\Delta \phi\,.
\end{equation}
Then we have:
\begin{Corollary}\ZLA{lemma:regol}  Let $ M(t)\in C^1(0,+\infty) $.
For every function $ f\in L^2(0,T;L^2(\Gamma) )$ and for every initial condition $ \xi\in L^2(\Omega) $, the function $ t\mapsto A^{-1} w^{f,\xi} (\cdot,t)$ is continuous from $ [0,+\infty) $ to $ L^2(\Omega) $.
\end{Corollary}
Thanks to this result, the following definition makes sense:

\begin{Definition}
We say that the initial condition $ \xi $ is controllable to $0  $ at time $ T $ if there exists $ f\in L^2(0,T;L^2(\Gamma)) $ such that $ A^{-1}w ^{f,\xi}(\cdot;T)=0\in L^2(\Omega)$.

We say that (CGM) is null controllable at time $ T $ if for every $ \xi\in L^2(\Omega) $ there exists $ f\in L^2(0,T;L^2(\Gamma)) $ such that $ A^{-1}w ^{f,\xi}(\cdot;T)=0\in L^2(\Omega)$. 
\end{Definition}

In the memoryless case, $ M(t)\equiv 0 $, the system is null controllable  at  \emph{any}
 time $ T>0 $. 
When $ M(t)\not= 0 $ but $ M(t)=0 $ for $ 0\leq t\leq T_0 $ then Eq.~(\ref{eq:CGM}) for $ t\leq T_0  $ coincide with the memoryless heat equation $ w_t=\alpha\Delta w $ and any initial condition can be controlled to $ 0 $ at any time $ T<T_0 $. 
Keeping  this fact in mind, our main result is: 
 
\begin{Theorem}\ZLA{Theo:mainNEGres}
Let $ \alpha>0 $ and let $ M(t)\in C^1(0,T) $, not identically zero. 
Let $ T $ be any time such that $ R(T)\neq 0 $, where $ R(t) $ is the resolvent kernel of $ M(t) $. 

There exist  initial data $ \xi $ which cannot be controlled to $ 0 $ at time $ T $.
\end{Theorem}

\subsection{Comments and references}

Under smoothness assumption on the kernel $ M(t) $, when $ \alpha=0 $ and $ M(0)>0 $, Eq.~(\ref{eq:CGM})   can be seen as a perturbation of the wave equation and its
  properties   resemble those of the wave equation.     In particular, the solutions belong to $ C(0,+\infty;L^2(\Omega)) $ for every $ f\in L^2_{\rm loc}(0,T;L^2(\Gamma)) $ and every initial condition $ \xi\in L^2(\Omega) $. Furthermore, there exists  $ T $  such that the reachable set 
\[ 
\left \{ w^{f,0}(\cdot,T)\,,\quad f\in L^2(0,T;L^2(\Gamma))\right \}
 \]
 is equal to $ L^2(\Omega) $. Several different techniques have been used in the proof, but the basic idea is always to compare with the wave equation, 
 see~\cite{AvdBeliski,Kim,PandLoretiSforza,PandAMO,Pandbeijing}. Furthermore, the infimum of the  control times is the same as that for the (memoryless) wave equation (see~\cite{AvdPAND,FU,Kim,PandIEOT,PandSHARP}).
 
 Instead, when $ \zaa>0 $ the properties of Eq.~(\ref{eq:CGM}) strongly resemble those of the standard, memoryless, heat equation in spite that   it is not possible to control an initial condition to be identically zero for every $ t>T $, where $ T $ is a preassigned time, see~\cite{IvanovPandolfi}. { So, it is a natural conjecture that     the controllability properties of system~(\ref{eq:CGM}) with $ \zaa>0 $ should be similar to those of the (memoryless) heat equation.  
 Along this line of thought,  it was proved 
 in~\cite{BarbuIannelli} that, for a suitable class of completely monotonic kernels, the reachable states at every time $ T>0 $ are dense in $ L^2(\Omega) $ and this supports the  conjecture that every initial condition $ \xi\in L^2(\Omega) $ can be controlled to hit the target $ \xi_0(x)\equiv 0 $ at a certain time $ T $, of course without remaining equal to zero in the future, due to the negative results in~\cite{IvanovPandolfi}.  
 }
 Hence,   the following negative fact was a surprise: there 
 { {\em exist\/}
 } kernels  $ M(t) $ which are even $ C^{\infty} $, and such that for every $ T>0 $ there  exist initial data which cannot be controlled to hit $ 0 $, see~\cite{Imanuv,HalanayPandolfi}. The proofs in these papers exibits two particular counterexamples.

   The goal of this paper is   the proof that in the presence of memory, i.e. for {\em every\/} smooth kernel $ M(t) $   not identically zero, there exist initial conditions which cannot be controlled to zero, as stated in Theorem~\ref{Theo:mainNEGres}.

\section{\ZLA{sect:PreliDEFIN}Preliminaries}
The number $ \alpha $ has to be positive and so, changing the time scale, i.e. replacing  $w( x,t)$ with $w( x,rt)$, we can assume  
\[ 
\alpha=1\,.
 \]
We present a transformation which simplifies the computations in this paper.   We consider a Volterra integral equation on $ t\geq 0 $
\[ 
y(t)+\intt M(t-s)y(s)\ZD s=f(t)\,.
 \]

 It is known (see~\cite[Ch.~2]{Gripe}) that it is uniquely solvable for every square integrable $ f(t) $, and that
 the solution is given by
 \[ 
y(t)=f(t)-\intt R(t-s) f(s)\ZD s\,. 
  \]
  The function $ R(t) $, the resolvent kernel of $ M(t) $, solves
  \[ 
  R(t)=M(t)-\intt M(t-s)R(s)\ZD s\,.
   \]
  We apply formally this transformation, ``solving'' Eq.~(\ref{eq:CGM}) with respect to the ``unknown'' $ \Delta w  $. We get
  \[ 
w _t=\Delta w +\intt R(t-s)w_s(s)\ZD s\,.
   \]
   Integrating by parts we get
  \begin{equation}
\ZLA{eq:CMM-modif}
w _t=\Delta w +a w(t)+\intt L(t-s)w(s)\ZD s-R(t)\xi\,,\qquad w(0)=\xi\,.
\end{equation}
Here,
\[ 
a=R(0)=M(0)\,,\qquad L(t)=R'(t)\,.
 \]
 
By, definition, a solution of Eq.~(\ref{eq:CGM}) is a solution of the Volterra integro-differential equation~(\ref{eq:CMM-modif}) (solutions can be defined in several different but equivalent ways).

Let us consider the operator $ A $ defined in~(\ref{eq:defiOPERa}).
It  is a selfadjoint operator with compact resolvent, which generates a holomorphic semigroup $ e^{At} $.

Let $ D $ be the ``Dirichlet operator'',
\[ 
u=Df\ \iff \mbox{$ u $ solves $ \left\{\begin{array}
{lll}
\Delta u=0 &{\rm in} \ \Omega\\
u=f &{\rm on} \ \Gamma=\partial \Omega\,.
\end{array}\right. $    }
 \]
A known fact (see~\cite[p.~180]{LAStriENCYCL}) is   the following:
\begin{Theorem}
\ZLA{teo:RegolarEQcalore}
Let $ f\in L^2(0,T;L^2(\Gamma)) $ and $ \xi\in L^2(\Omega) $.
 The solution to the heat equation
\[ 
\zthe _t=\Delta\zthe   +g\,,\qquad \zthe   (x,0)=\xi(x)\,,\quad \zthe   (x,t)=f(x,t)\ x\in\partial\Omega
 \]
 is given by
 \begin{equation} \ZLA{eq:solocalore}
 \zthe   (\cdot,t)= \zthe^{f,\xi,g}   (\cdot,t)=e^{At}\xi+\intt e^{A(t-s)}g(s)\ZD s-A\intt e^{A(t-s) } Df(s)\ZD s\,.
  \end{equation}
The solution   is unique in $ L_{\rm loc}^2(0,+\infty;L^2(\Omega)) $ and $A^{-1}\theta(\cdot,t)\in C(0,+\infty;L^2(\Omega)) $. Furthermore, if $ \xi=0 $ then $ \zthe(\cdot,t)\in L^2_{\rm loc}(0,+\infty,H^{1/2}(\Omega) )$.
 \end{Theorem}
 We apply formula~(\ref{eq:solocalore}) to~(\ref{eq:CMM-modif}) with
 \[ 
 g(t)=a w(t)+\intt L(t-s)w(s)\ZD s-R(t)\xi
  \]
  and we find the following Volterra integral equation for $ w(x,t) $:
 
\begin{eqnarray}
\nonumber &&w(x,t)-\intt e^{A(t-s)}\left [
aw(s)+\ints L(s-r) w(r)\ZD r
\right ]\ZD s\\
&&
\ZLA{eq:integralEQperSOLUZ} =\left \{ e^{At}\xi-\intt e^{A(t-s)}R(s)\xi\ZD s\right \}-A\intt e^{A(t-s)}Df(s)\ZD s
\end{eqnarray}
  
 Theorem~\ref{teo:propreSOL} and Corollary~\ref{lemma:regol}   follow  from this formula, thanks to the properties of the solutions of the (memoryless) heat  equation stated in Theorem~\ref{teo:RegolarEQcalore}.
 
 See~\cite{Laksh} for the theory of   Volterra integral and integro-differential equations in Banach spaces, and~\cite{DaPrato} for further information on the semigroup approach to boundary value problems for parabolic equations.

   \subsection{Projection of the system on the eigenspaces}
   
   The previous results  allows us to project the system on the eigenvectors of the operator  $ A $.
Let $\{ \phi_n\}   $ be an orthonormal basis of $ L^2(\Omega) $, whose elements are   eigenvectors  of the operator $ A $ in~(\ref{eq:defiOPERa}). So we have:
\[ 
\Delta \phin=-\zln\phin\,,\qquad \phin(x)=0\  {\rm on}\   \Gamma= \partial\Omega\,.
 \]
Note that $ \zln>0 $.

Let
\[ 
\wn(t)=\int _{\Omega}w(x,t)\phin(x)\ZD x\qquad \xi_n=\int _{\Omega}\xi(x)\phin(x)\ZD x\,.
 \]
Then $ \wn(t) $ solves
\[ 
\wn'(t)=\an\wn+\intt L(t-s)\wn(s)\ZD s-R(t)\xi_n- g_n(t) 
 \]
 where 
 \[ 
 g_n(t)= \int _{\Gamma} (\gamma_1\phin  )f(x,t)\ZD\Gamma
  \]
  ($\gamma_1   $ denotes normal derivative on $ \Gamma $) and
  \begin{equation}
\ZLA{eq:rappreW}w(x,t)=\sum \phi_n(x) w_n(t)\,.
\end{equation}

We introduce
\[ 
\mun=\zln-a 
 \]
(we have  $ \mu_n>0 $ for  large $ n $) 
so that
\begin{eqnarray}
\nonumber 
&&w_n(t)-\intt e^{-\mun (t-\tau)}\int_0^{\tau} L(\tau-s) w_n(s)\ZD s\,\ZD\tau\\
\ZLA{eq:rappreWn}&&=\left (e^{-\mun t} -\intt e^{-\mun (t-s)}R(s)\ZD s\right )\xi_n -\intt e^{-\mun (t-s)} g_n(s)\ZD s\,.
\end{eqnarray}

Let 
$ T>0 $. We define a transformation 
$ \cal L $ in $ L^2(0,T;L^2(\Omega)) $, as follows: 
\[ 
{\cal L}\left (\sum \phin(x) h_n(t)\right )=\sum \phin(x)\left ({\cal L}_n h_n\right )(t)
 \]
 where 
 \[ 
 \left ({\cal L}_n h\right )(t)=\intt e^{-\mun (t-s)}\int_0^s L(s-r) h(r)\ZD r\,\ZD s\,.
  \]
  Then we have
  \begin{align}
\nonumber &
\left (I-{\cal L}\right )w=\sum \phin(x)\left \{
\left (e^{-\mun t} -\intt e^{-\mun (t-s)}R(s)\ZD s\right )\xi_n \right.
\\&
\ZLA{eq:semiexplicW}\left.-\intt e^{-\mun (t-s)} g_n(s)\ZD s
\right \}\,.
\end{align}
We prove:

 \begin{Lemma}\label{Lemma:ContLn}
The transformation $ \cal L $ in $ L^2(0,T;L^2(\Omega)) $ is linear and continuous. The transformation $ (I-{\cal L}) $ is invertible and its inverse is continuous.
\end{Lemma}
\zProof
Linearity  is clear. We prove the continuity of $ {\cal  L} $ and of its inverse, using the fact that $ \{\phin\} $ is an orthonormal basis of $ L^2(\Omega) $. This implies that
 \[
 \left \| \left (\sum h_n(t)\phin(x)\right )\right \|^2  _{L^2(0,T;L^2(\Omega)) }
 =\sum \int_0^T |h_n(t)|^2\ZD t\,.
\]
Then we have:
\begin{align*}&\int_0^T \left |
\left ({\cal L}_n h\right )(t)
\right |^2\ZD t
=\int_0^T \left |\int_0^t
e^{-\mun (t-s)}\int_0^s L(s-r)h(r)\ZD r\,\ZD s
\right |^2\ZD t\\ 
&\leq T^2  \left ( \int_0^T e^{-2  \mun s} \ZD s \right ) 
  \left (
\int_0^T L^2(s)\ZD s
\right )\left (  \int_0^T h^2(r)\ZD r\right )\ZD s
\\
&\leq C \int_0^T |h(s)|^2\ZD s\,.
  \end{align*}
We can chose the constant $ C $ independent of $ n $ thanks to the fact that $ \mun >0 $ for large $ n $.
So, we have
\begin{align*}&  \left \| {\cal L}\left (\sum h_n(t)\phin(x)\right )\right \|^2  _{L^2(0,T;L^2(\Omega)) }=
\int_0^T \int_\Omega \left |
\sum \left ( \mathcal{L}_ n h_n  \right  )(t)\phi_n(x)
\right |^2\ZD x\,\ZD t\\
&=
\int_0^T \sum \left |
\left (
{\cal L}_n h_n
\right )(t)
\right |^2\ZD t 
 \leq C\sum
\int_0^T |h_n(s)|^2\ZD s\\
&=  C  \left \|\left (\sum h_n(t)\phin(x)\right )\right \|^2  _{L^2(0,T;L^2(\Omega)) }
\,.
\end{align*}
This proves continuity of the transformation $ \cal L $ and so  also of $ I-{\cal L}  $.  In order to prove that this last transformation  has a bounded inverse, we exibit explicitly its inverse.

To compute the inverse, we must solve, for every $ k(x,t)=\sum \phi_n(x)k_n(t) $,
\[ 
\left (I-{\cal L} \right )\left (\sum \phin(x) f_n(t)  \right )= k(x,t)=\sum \phi_n(x)k_n(t)
 \]
 i.e.
\[
 \sum \phin \left \{
f_n(t)-\intt f_n(\tau)\int_0^{t-\tau} L(t-\tau-s) e^{-\mun s}\ZD s\,\ZD\tau
\right \} 
 =\sum \phin(x) k_n(t)\,.
\]
We introduce $ H_n(t) $, the resolvent kernels of 
\begin{equation}\ZLA{eq:defiZn}
Z_n(t)=- \int_0^{t} L(t-s) e^{-\mun s}\ZD s\,. 
 \end{equation}
Then we must choose
\[ 
f_n(t)=  
k_n(t)-\intt H_n(t-s) k_n(s)\ZD s
 \]
and so
\[ 
\left (I-
{\cal L} 
\right )^{-1} \sum \phin(x) k_n(t)=\sum\phin(x)\left \{
k_n(t)-\intt H_n(t-s)k_n(s)\ZD s
\right \}\,.
 \]
Continuity of this transformation is seen as above,  using 
the fact that $ \mun>0 $ for large $ n $, so that
  $ |Z_n(t)|\leq M/\mun $ (for large $ n $) where $ M=M_T $. So,   Gronwall inequality applied to
 \[ 
 \left | H_n(t)\right | \leq \left | Z_n(t)\right |+\intt \left |Z_n(s)\right |\cdot \left |H_n(s)\right |\ZD s
  \]
   gives
  \[ 
 \left | H_n(t)\right |\leq \frac{M}{\mun}\,,\qquad M=M_T\,. 
   \]
Continuity now follows as above.\zdia

Using~(\ref{eq:semiexplicW}) we find that
\begin{align}
&\nonumber    
 w(x,t)  =\left (I-{\cal L}\right )^{-1}\sum \phin(x)\left \{
\left (e^{-\mun t} 
  -\intt e^{-\mun (t-s)}R(s)\ZD s\right )\xi_n\right. 
  \\\nonumber& 
   \left. -\intt e^{-\mun (t-s)} g_n(s)\ZD s
\right \} 
 =\sum\phin(x)\left \{-\left [\intt e^{-\mun (t-s)}g_n(s)\ZD s\right.\right.
 \\ \nonumber &
 \left.\left. +\intt H_n(t-\tau)\int_0^{\tau} e^{-\mun (\tau-s)}g_n(s)\ZD s\,\ZD\tau\right ]\right .
 \\ 
 \nonumber&
 \left . +\left [
e^{-\mun t}-\intt e^{-\mun(t-s)}  R(s)\ZD s \right.\right.
 \\&
 \left.\left.
-\intt H_n(t-\tau)
\left (
e^{-\mun \tau} -\int_0^\tau e^{-\mun (\tau-s)} R(s)\ZD s
\right )\ZD\tau
\right ]
\xi_n\right\}
\ZLA{eq:explicitW}
\end{align}

Now we recall the definition of controllability at time $ T $ and we can state:

\begin{Theorem}\ZLA{teo:equivCONTROmoment}
Equation~(\ref{eq:CGM}) is controllable to $ 0 $ at time $ T $ if for every sequence $ \{\xi_n\}\in l^2 $ there  exists a function $ f\in L^2(0,T;L^2(\Gamma)) $ which solves the following moment problem:
\begin{align}
\nonumber& \left [\intT e^{-\mun (T-s)}
\left (\int _{\Gamma} (\gamma_1\phin )f(x,s)\ZD\Gamma\right )
\ZD s\right.\\
\nonumber & \left.-\intT H_n(T-\tau)\int_0^{\tau} e^{-\mun (\tau-s)}
\left (
\int _{\Gamma} (\gamma_1\phin )f(x,s)\ZD\Gamma
\right )\ZD s\,\ZD\tau\right ]\\
\nonumber&  =\left [
e^{-\mun T}-\intT e^{-\mun( T-s)}  R(s)\ZD s\right.\\
& \left. -\intT H_n(T-\tau)
\left (
e^{-\mun \tau}-\int_0^\tau e^{-\mun (\tau-s)} R(s)\ZD s
\right )\ZD\tau
\right ]
\xi_n
\ZLA{eq:MOMprob}
\end{align}
\end{Theorem}
The proof of Theorem~\ref{Theo:mainNEGres} is then reduced to the proof that this moment problem is not solvable.
\section{The proof of Theorem~\ref{Theo:mainNEGres}}

Let $ N_0 $ be such that
\[ 
n\geq N_0\ \implies \mun>0\,.
 \]
 We shall consider the moment problem in Theorem~\ref{teo:equivCONTROmoment} only for the indices $ n\geq N_0 $ and we shall prove that it can't be solved.
  
We first examine the right hand side of~(\ref{eq:MOMprob}). We recall that $ H_n(t) $  is the resolvent kernel of $ Z_n(t) $ in~(\ref{eq:defiZn})
 so that the following equality holds:
 \[ 
 H_n(t)=-\intt L(t-s) e^{-\mun s}\ZD s+\intt \left [
 \int_0^{t-\tau} L(t-\tau-s) e^{-\mun s}\ZD s
 \right ]H_n(\tau)\ZD\tau
  \]
The function $ L(t) $ is bounded on $ [0,T] $ for every $ T>0 $ and $ \mun>0 $, so, using Gronwall inequality,  there exists $ C $ (which depends on $ T $ but not on $ n $) such that
\[ 
\left | H_n(t)\right |\leq C\frac{1}{\mun } 
 \]
(a fact already used in the proof of Lemma~\ref{Lemma:ContLn}). 

We fix $ T$ such that $ R(T)\neq 0 $.
On every compact interval, using boundedness of $ M'(t) $   hence of $ R'(t) $, we have:
 \begin{align*}
 & \int_0^T R(s) e^{-\mun (T-s)}\ZD s=\frac{1}{\mun }\left ( R(T)-e^{-\mun T} R(0)-\int_0^T e^{-\mun (T-s)}R'(s)\ZD s\right )\,,\\
 & \left |\int_0^T e^{-\mun (T-s)}R'(s)\ZD s\right | \leq \frac{{\rm const}}{\mun}\,,\\
 & \left | \int_0^T H_n(T-\tau) \left ( e^{-\mun \tau}+\int_0^\tau e^{\mun (\tau-s) }R(s)\ZD s\right )\ZD\tau \right |\leq \frac{{\rm const}}{\mun}\,.
  \end{align*}
 
 Let
 \begin{align*}
&d_n= \left [
e^{-\mun T}-\int_0^T e^{-\mun(T-s)}  R(s)\ZD s\right. \\  
& \left.-\int_0^T H_n(T-\tau)
\left (
e^{-\mun \tau}-\int_0^\tau e^{-\mun (\tau-s)} R(s)\ZD s
\right )\ZD\tau
\right ]
\xi_n
\\
&=\left [
e^{-\mun T} - \frac{1}{\mun}\left (R(T) -e^{-\mu_n^2 T} R(0)-\int_0^T e^{-\mu_n^2(T-s)} R'(s)\ZD s
\right)\right.\\
&\left.-\int_0^T H_n(T-\tau)\left ( e^{-\mu_n^2\tau} -\int_0^{\tau} e^{-\mu_n^2(\tau-s)} R(s)\ZD s\right )\ZD\tau
\right ]\xi_n
 \,.
  \end{align*}
  
  Using the existence of $ C $ such that
  \[ 
  \mun e^{-\mun T}< \frac{C}{\mun}
   \]
 the previous equalities, with $ R(T)\neq 0 $, give
 \[ 
 \mun d_n=\left (- R(T)+\frac{M_n}{\mun}\right )\xi_n
  \]
  where $ \{M_n\} $ is a \emph{bounded} sequence. Hence, we get:
  \begin{Lemma}
   Let $ R(T)\neq 0 $.   There exists $ N>N_0 $ with the following property: for every $ \{c_n\}\in l^2([N,+\ZIN)) $ the equation in $l^2([N,+\ZIN)) $
  \[ 
  \mun d_n=\left (- R (T)+\frac{M_n}{\mun}\right )\xi_n=c_n
   \]
   admits a solution $ \{\xi_n\}\in l^2([N,+\ZIN))  $.
   \end{Lemma}
   
   We go back to the moment problem~(\ref{eq:MOMprob}) for $ n\geq N $. If equation~(\ref{eq:CGM}) is controllable to $ 0 $ at time $ T $, then the moment problem
   \begin{align}
\ZLA{eq:NEWmoPRO}
\nonumber& \left [\intT e^{-\mun ( T-s)}
\left (\int _{\Gamma} \mun(\gamma_1\phin   )f(x,s)\ZD\Gamma\right )
\ZD s\right.\\
\nonumber & \left.-\intT H_n(T-\tau)\int_0^{\tau} e^{-\mun (\tau-s)}
\left (
\int _{\Gamma} \mun(  \gamma_1\phin )f(x,s)\ZD\Gamma
\right )\ZD s\,\ZD\tau\right ]=c_n
\end{align}
is solvable for every   sequence $ \{c_n\}\in l^2=l^2(N,+\infty) $.
We exchange the order of integration and we rewrite this equalities as
\begin{equation}\ZLA{eqMOMEproFIN}
\int_0^T \int_\Gamma f(x,T-s) E_n(x,s)
\ZD\Gamma\,\ZD s=c_n \,,\qquad n\geq N
 \end{equation}
where
 \[ 
 E_n(x,s)=\mun (\gamma_1 \phin )\left (
e^{-\mun s}-\int_0^s H_n(s-\tau)e^{-\mun \tau}\ZD\tau
\right )\in L^2(0,T;L^2(\Gamma))\,.
  \]
   We recall from~\cite[Theorem~I.2.1]{AvdoninIVANOV} that the moment problem~(\ref{eqMOMEproFIN}) is solvable for every $ l^2 $-sequence $ \{c_n\} $ ($ n\geq N $) if and only if the sequence $ \{ E_n(x,t)\} $ admits a \emph{bounded } biorthogonal sequence $ \{\chi_n(x,t)\} $ in { $L^2(0,T;L^2(\Gamma))$}; i.e. if and only if there exists a \emph{bounded } sequence $ \{\chi_n(x,t)\} $ in {$ L^2(0,T;L^2(\Gamma))$} such that
   \[ 
   \int_0^T\int_\Gamma E_n(x,t)\chi_k(x,t)\ZD \Gamma\,\ZD t=\delta_{n,k}=\left\{
   \begin{array}{lll}
   1 &{\rm if}& n=k\\
    0 &{\rm if}& n\neq k\,.
   \end{array}
   \right.
    \]
   
   We are going to prove that this sequence does not exist, relaying on known properties of the (memoryless) heat equation. We proceed in two steps:  the first step computes ``explicitly''   $H_n(t)$. The second step, using this expression of $ H_n(t) $, shows that  {\em a bounded sequence $ \{\chi_n(x,t)\} $ does not exist, i.e. the moment problem is not solvable.\/}
   
   We proceed with the proof.
   
   \subparagraph{Step 1: a formula for $ H_n(t) $.}
   
   Here we find a  formula for $ H_n(t) $,  for every \emph{fixed } index $ n $. So, for clarity, the fixed index $ n $ is not indicated in the  computations and $ H_n(t) $ (any fixed $ n $) is denoted $ H(t) $. Analogously, $ \mun $, with $ n $ fixed, is indicated as $ \mu^2 $. Furthermore, we use $ \star $ to denote  the convolution,
   \[ 
   f\star g=(f\star g)(t) =\intt f(t-s)g(s)\ZD s\,.
    \]
    We shall use the commutativity and the associativity   of the convolution:
    \[ 
    f\star g=g\star f\,,\qquad f\star(g\star h)=f\star(g\star h)\,.
     \]

     The convolution of a function with itself is denoted as follows:
      \[ 
  f^{ \star 1 }=f\,,\qquad      f^{ \star 2 }=f\star f\,,\quad  f^{ \star k }=f\star  f^{ \star( k-1) }\,.
       \]
  Let
    \[ 
    e_k(t)=\frac{t^k}{k!}e^{-\mu^2 t}\quad \mbox{so that} \quad  e_0\star e_k=e_{k+1}\,.   
     \]
  
   By definition, $ H(t) $ is the resolvent kernel of
   \[ 
   Z(t)=-\intt L(t-s) e^{-\mu^2 s}\ZD s=-L\star e_0\,.
    \]
  
 We shall use:
    
  \begin{Lemma}
    Let    $ G(t) $ be any (integrable) function and $ \tilde G=G\star e_k $. Then,
    \[ 
 Z\star \tilde G=e_{k+1}   \star (-L\star G)
     \]
\end{Lemma}
 In fact:    
     \[ 
     Z\star \tilde G=(-L\star e_0)\star(G\star e_k)=(e_0\star e_k)\star (-L\star G)=e_{k+1}\star (-L\star G)\,.
      \]

       The previous lemma shows that
       \[ 
      Z^{ \star k } =(-1)^kL^{ \star k }\star  e_{k-1}\,.
        \]
        The known formula of the resolvent (\cite[p.~36]{Gripe}) gives
       
  \begin{align} 
    \nonumber   H(t)=\sum _{k=1}^{+\infty} 
      (-1)^{k-1} Z^ {\star k }
       =-\sum _{k=1} ^{+\ZIN}L^{ \star k }\star e_{k-1}=\\
            \ZLA{eq:LaDefiH}
 =-      \int_0^t \left (\sum _{k=1} ^{+\ZIN}L^{ \star k }(t-s)
\frac{s^{k-1}}{(k-1)!}    \right )e^{-\mu^2 s}\ZD s\,. 
\end{align}
The series converges uniformly since  the following holds:
\[ 
|  L(t)|<M \quad 0\leq t\leq T\ \implies |L^{ \star k }(t)|\leq \frac{T^k M^k}{k!}\quad 0\leq t\leq T\,.
 \]

\subparagraph{Step 2: the bounded biorthogonal sequence does not exist.}
We reintroduce dependence on the index $ n $. So 
\[ 
e_k(t)=\frac{t^k}{k!}e^{-\mun t}\,.
 \]

We go back to the moment problem~(\ref{eqMOMEproFIN}). We prove that
 it is not solvable as follows: we prove that  if the sequence $ \{E_n(x,t)\} $ admits a biorthogonal sequence $ \{\chi_k(x,t)\} $, then this sequence cannot be bounded. So, let

\begin{equation}\ZLA{eq:FineBio}
\delta_{n,k}=\mun\int_\Gamma (\gamma_1 \phi_n(x) ) \left [\int_0^T
\chi_k(x,t)\left (
e^{-\mun t}-\intt H_n(t-\tau)e^{-\mun\tau}\ZD\tau
\right )\ZD t
\right ]\ZD\Gamma\,.
 \end{equation}
We have, using~(\ref{eq:LaDefiH}): 

 \begin{align*}
& \intt H_n(t-\tau)e^{-\mun \tau}\ZD\tau=	e_0\star H_n=-e_0\star \left (
 		\sum _{k=1}^{+\infty} L^{(\star k)}\star e_{k-1}\right )\\
 		&=-\sum _{k=1}^{+\infty} L^{(\star k)}\star e_{k } 
 	 =-\intt\left [ \sum _{k=1}^{+\ZIN} L^{(\star k)}(t-s)\frac{s^{k }}{ k !}\right ] e^{-\mun s}\ZD s
 		=\intt G(t,s) e^{-\mun s}\ZD s\,.
  \end{align*}
  
  Note that $ G(t,s) $ does not depend on $ n $ and
  equality~(\ref{eq:FineBio}) can be written as
\begin{align*} 
&\delta_{n,k}=
  \int_\Gamma\intT(\gamma_1\phi_n(x))\left( \mun e^{-\mun r}\right )\left [
\chi_k(x,r)-\int _r^{T} G(s,r) \chi_k(x,s)\ZD s
\right ]\ZD r\,\ZD\Gamma\\
&=
\intT\left( \mun e^{-\mun r}\right )\left \{ \int_\Gamma (\gamma_1\phi_n(x)) \left [
\chi_k(x,r)-\int _r^{T} G(s,r) \chi_k(x,s)\ZD s
\right ] \ZD\Gamma\right \} \ZD r\,.
 \end{align*} 
 Hence, the sequence $ \{\tilde \Psi_k(r)\} $,
 \[ 
 \tilde\Psi_k(r)= \int_\Gamma (\gamma_1\phi_n(x)) \left [
\chi_k(x,r)-\int _r^{T} G(s,r) \chi_k(x,s)\ZD s
\right ] \ZD\Gamma\,,\quad n\geq N
  \]
  is biorthogonal to  
 $ \{   \mun e^{-\mun r}\} $ in $ L^2(0,T) $.
 
 Now we invoke the following result from~\cite{TAOcorrect}:
 \begin{Lemma}
There exist two positive numbers $ m $ and $ M $ such that the following holds for every index $ n $:
\[ 
0<m\leq \int_\Gamma \left |\frac{\gamma_1\phi_n(x)}{\n}\right |^2\ZD \Gamma\leq M\,.
 \]
\end{Lemma}

Consequently, the sequence
\[ 
\{\Psi_k(t)\}\,,\qquad \Psi_k(t)=\frac{1}{\n}\tilde\Psi_k(t) 
 \]
 is a \emph{bounded} biorthogonal sequence of $ \{\mun \n e^{-\mun t}\} $ in $ L^2(0,T) $. 
 
 We proved in~\cite{HalanayPandolfi} that for every $ T>0 $ the sequence $ \{\mun \n e^{-\mun t}\} $ does not admit any bounded biorthogonal sequence in $ L^2(0,T) $  and {\em this completes the proof of Theorem~\ref{Theo:mainNEGres}.\/}
 
 For completeness, we sketch the proof of this last statement (see~\cite{HalanayPandolfi} for additional   details):
 \begin{Lemma}
Any sequence $ \{\Psi_n(t)\} $ which is biorthogonal to $ \{\mun \n e^{\mun t}\} $ in $ L^2(0,T) $ is unbounded.
\end{Lemma}
\zProof Let 
$e_n$ be the function $e^{-\mun t}$ in $L^2 (0,\infty)$ and  denote  by $e^T_n$ its restriction to $(0,T)$.
\[
E(\infty)={\rm cl\, span}\{e_n\} \subseteq L^2 (0,\infty)\,,\qquad
E(T)={\rm cl\, span}\{e^T_n\} \subseteq L^2 (0,T)
\,.
\] 
$ E(\infty) $ is a proper subspace of $ L^2 (0,\infty) $ 
(M\"untz Theorem, see~\cite{Schw}) . Let $P_T: L^2 (0,\infty) \to L^2 (0,T)$ be the   operator $P_T f = f|_{(0,T)}$. The operator $P_T$ is an isomorphism between $E(\infty)$ and $E(T)$ (see~\cite[formula~(9.a) p.~55]{Schw}).

Suppose that $\{\tilde \psi_n\} $ is any biorthogonal to $\{e^T_n\} $ in $L^2 (0,T)$.
We prove  that the sequence $\{\tilde \psi_n\} $ is   {\em exponentially unbounded.\/}

Let $\psi_n $ be the orthogonal projection of $ \tilde \psi_n $ on $E(T)$. Then, $ \{\psi_n\} $ is biorthogonal to $ \{ e_n^T\} $ too and
\[ 
\| \psi_n\|_{L^2(0,T)}\leq \|\tilde \psi_n\|_{L^2(0,T)}\,.
 \]

We have { ( $ (\cdot,\cdot) $ is the inner product in the indicated spaces)  }

\[
\delta_{jn} = (\psi_j, e^T_n)_{L^2(0,T)}= (\psi_j, e^T_n)_{E(T)}=  (\psi_j, P_T e_n)_{E(T)} = (P^*_T \psi_j, e_n)_{E(\infty)}\,.
\]
 Hence  $\{P^*_T \psi_n\} $ is biorthogonal  to $\{e_n\} $ and furthermore $\varphi_n=P^*_T \psi_n\in E(\infty)$  since   $ P_T\in{\cal L}( E(\infty),E(T))$.
 Hence, $ \{\varphi_n\} $ is the   biorthogonal sequence of $ \{ e_n\} $ whose $ L^2 $-norm is minimal.
 
Using~\cite[Lemma~3.1]{F-R} we have:

\begin{equation}\label{eq:1.8}
||\varphi_n||_{L^2 (0,\infty)} = \frac{2}{n^2} e^{[\pi + O(1)]n}, \quad n \to \infty\,. 
\end{equation}

Since $P^*_T    \in{\cal L}( E(T) , E(\infty))$ is  boundedly invertible,  there exist \emph{positive numbers} $m $ and  $M  $ such that for every $ n $ we have
$$
m\|\psi_n\|_{L^2 (0,T)} \le \|P^*_T \psi_n \|_{L^2(0,+\infty)}   \le M \|\psi_n\|_{L^2 (0,T)} 
$$
  since  $  P^*_T \psi_n   = \varphi_n  $.
It follows that
\begin{equation}
\label{eq:1.9}
\|\tilde\psi_n\|_{L^2 (0,T)}   \geq\|\psi_n\|_{L^2 (0,T)} \ge \frac{1}{M} \|\varphi_n\|_{L^2 (0,\infty)} \quad \forall n\,. 
\end{equation}
So,  {\em any\/} biorthogonal sequence of $\{e^{-\mun t}\} $ in $ L^2(0,+\infty) $ is {\em exponentially\/} unbounded and from~(\ref{eq:1.9}) we see that       {\em any\/} biorthogonal sequence of $\{  e^{-\mun t}\}_{n\ge N_T}$ in $ L^2(0,T) $ is {\em exponentially\/} unbounded too.\zdia

  Let's go  back to the sequence $ \{\Psi_n(t)\} $. This sequence cannot be bounded. Otherwise,  
the sequence $ \{\mun\n \Psi_k(t)\} $ is a biorthogonal sequence to $\{ e^{-\mun t}\} $ 
such that
\[ 
\| \mun\n \Psi_k(t)\|_{L^2(0,T)}\leq  C \mun\n\leq n^{3/d}\,,\qquad d={\rm dim}\,\Omega,
 \]
 using known estimates on the eigenvalues of the laplacian   (see~\cite[p.~192]{mikailovLIBRO}).
 
 \begin{Remark} {\rm
 Instead of a time $ T $ in which $ R(T)\neq 0 $ we might have used a time $ T $ at which $ R^{(k)}(T)\neq 0 $ and $ R^{(m)}(T)=0 $ for $ m<k $, but this does not change the content of Theorem~\ref{Theo:mainNEGres} in an essential way.\zdia

 }
 \end{Remark}


\begin{thebibliography}{99}
  
  \bibitem{AvdBeliski}   S. Avdonin, B.P.  Belinskiy, On controllability of an 
homogeneous string with memory, J. Mathematical Analysis Appl., { 398} (2013) 254--269.

 \bibitem{AvdoninIVANOV} 
  S. A. Avdonin, S.A. Ivanov, {\em Families of exponentials. The Method of Moments in Controllability Problems for Distributed Parameter Systems.\/}  Cambridge University Press, New-York,~1995.
  
  \bibitem{AvdPAND} S. Avdonin and  L. Pandolfi,
 {Simultaneous temperature and flux controllability for heat equations with memory},
  {\em Quarterly Appl. Math.,\/} Electronic version DOI~10.1090/S0033-569X-2012-01287-7
  
 \bibitem{BarbuIannelli}  V. Barbu, M. Iannelli, 
 {Controllability of the heat equation with
  memory.}  { \em  Diff. Integral Eq.\/}
   {  13}~(2000)      1393-1412.
   
   \bibitem{DaPrato}
   A. Bensoussan,   G. Da Prato,  M.C. Delfour,  S.K. Mitter,   {\em Representation and control of infinite dimensional systems.\/}   Birkh\"auser  Boston, MA, 2007.
%
\bibitem{ColemGURTIN} B.D. Colemann, M.E. Gurtin, Equipresence and constitutive equations for heat conductors. {\em Z. Angew. Math. Phys.\/}  18~(1967) 199-208.

\bibitem{F-R} H.O. Fattorini, D.L. Russell, {  Exact Controllability Theorems for Linear Parabolic Equations in One Space Dimension.}  {\em Arch. Rational Mech. Anal.\/} 43~(1971)  272-292.

   \bibitem{FU} X. Fu, J. Yong, X. Zhang,
   {  Controllability and observability of the heat
equation with hyperbolic memory kernel.\/}
  {\em    J. Diff. Equations.\/}  { 247}~(2009)  2395-2439.
  
  
  \bibitem{Gripe} G. Gripenberg, S.-O.  Londen,  O. Staffans,  
{\em Volterra integral and functional equations.\/}
Encyclopedia of Mathematics and its Applications, 34. Cambridge University Press, Cambridge, 1990.
  
 
%
  \bibitem{Imanuv} S.Guerrero and O.Y.Imanuvilov, Remarks on non controllability of the heat equation with memory, {\em 
ESAIM: Control, Optimisation and Calculus of Variations ,\/}
{  19} (2013) 288-300.
%

 \bibitem{GurtinPipkin}  M.E. Gurtin, A.G. Pipkin,   A general theory of heat conduction with finite wave speed. { \em
Arch. Rat. Mech. Anal.\/}~31~(1968)  113-126.
  
  
   \bibitem{HalanayPandolfi}  A. Halanay, L. Pandolfi, Lack of controllability of the heat equation with memory, {\em Systems \& Control Letters,\/} { 61} (2012) 999-1002.
%
%
  
  
  
  
  \bibitem{TAOcorrect}A. Hassel, T. Tao,
  Upper and lower bounds for normal derivatives of Dirichlet eigenfunctions. {\em Math. Res. Lett.\/} 9 (2002),  289--305;  Erratum for "Upper and lower bounds for normal derivatives of Dirichlet eigenfunctions'', {\em Math. Res. Lett.\/} { 17} (2010), no. 4, 793--794.
   
   \bibitem{IvanovPandolfi}S. Ivanov, L. Pandolfi,  {  Heat equation with memory: Lack of controllability to rest.\/} { \em  J. Math. Anal. Appl.\/}~355~(2009) 1-11

\bibitem{Kim} J.U. Kim, Control of a second-order integro-differential equation. {\em SIAM  J. Control Optim. \/}~31~(1993) 101-110.
  
  
  \bibitem {LAStriENCYCL}   I. Lasiecka, R.Triggiani,     {\em Control theory for partial
differential equations: continuous and approximation theories. II.
Abstract hyperbolic-like systems over a finite time horizon.\/}
Encyclopedia of Mathematics and its Applications, 75. Cambridge
University Press, Cambridge, 2000.
  
   
  
   


  \bibitem{Laksh}  V. Lakshmikantham, R.M. Rao, \emph{Theory of integro-differential equations,\/} Gordon  \& Breach, Lausanne, 1995.
  
    \bibitem{Lions}  J.-L. Lions,   {\em  Contr\^ole  optimal de syst\`emes gouvern\'es par des  \'equations aux  d\'eriv\'ees  partielles.  \/} Dunod, Paris, 1968.
    
    
\bibitem{PandLoretiSforza}P. Loreti, L. Pandolfi, D. Sforza,  Boundary controllability and observability   of a viscoelastic string,    {\em SIAM J. Control Optim.\/}   50 (2012) 820-844.

\bibitem{mikailovLIBRO}  V. P. Mikhailov, {\em Partial Differential Equations,\/} Mir, Moscou 1978.

\bibitem{PandAMO} L. Pandolfi,
 {The controllability of the Gurtin-Pipkin equation: a cosine operator approach.} 
 {\em Appl.Math. and Optim.\/}~52~(2005)   143-165;  Erratum to: The controllability of the Gurtin-Pipkin equation: a cosine operator approach. {\em  Appl. Math. Optim.\/}~64~(2011)   467-468.
 
 \bibitem{PandIEOT} L. Pandolfi, Riesz systems and   controllability   of   heat equations with memory. {\em  Int. Eq. Operator Theory.\/}~64~(2009)  429-453.
 
\bibitem{Pandbeijing} L. Pandolfi, Riesz systems and moment method in the study of heat equations with memory  in one space dimension.  {\em Discr. 
Cont. Dynamical Systems,
Ser. B.\/}~14~(2010) 1487-1510.
 

\bibitem{PandSHARP} L. Pandolfi, Sharp control time in viscoelasticity, in preparation.

  \bibitem{Preziosi}  D.D. Joseph and L. Preziosi,
 { Heat waves,\/}
 { \em Rev. Modern Phys.,\/} {  61} (1989),  41--73;
  {  Addendum to the paper: ``Heat waves'',\/}
 { \em  Rev. Modern Phys.,\/} { 62} (1990),  375--391.
 
  \bibitem{Schw} L. Schwartz, {\em Etude des sommes d'exponentielles.}     Hermann, Paris 1959. 
  \end{thebibliography}
\end{document}